\documentclass[twocolumn,showpacs,preprintnumbers,amsmath,amssymb]{revtex4}
\usepackage{graphicx}
\newcommand{\be}{\begin{equation}}

\newcommand{\ee}{\end{equation}}
\newcommand{\bea}{\begin{narray}}
\newcommand{\eea}{\end{eqnarray}}
\newcommand{\beas}{\begin{eqnarray*}}
\newcommand{\eeas}{\end{eqnarray*}}

\begin{document}

\title{Interplay of initial deformation and Coulomb proximity on nuclear decay} 

\author{S. Hudan}
\author{R. Alfaro}
\author{L. Beaulieu}
\author{B. Davin}
\author{Y. Larochelle}
\author{T. Lefort}
\author{V.E. Viola}
\author{H. Xu}
\author{R. Yanez} 
\author{R.T. de Souza} 
\affiliation{
Department of Chemistry and Indiana University Cyclotron Facility \\ 
Indiana University, Bloomington, IN 47405} 

\author{R.J. Charity}
\author{L.G. Sobotka}
\affiliation{
Department of Chemistry, Washington University, St. Louis, MO 63130}

\author{T.X. Liu}
\author{X.D. Liu}
\author{W.G. Lynch}
\author{R. Shomin}
\author{W.P. Tan}
\author{M.B. Tsang} 
\author{A. Vander Molen} 
\author{A.Wagner}
\author{H.F. Xi}
\affiliation{
National Superconducting Cyclotron Laboratory and Department of
Physics and Astronomy \\ 
Michigan State University, East Lansing, MI 48824}

\date{\today}

\begin{abstract}

Alpha particles emitted from
an excited projectile-like fragment (PLF$^*$) formed in a peripheral collision of
two intermediate-energy heavy ions exhibit a strong preference for emission
towards the target-like fragment (TLF).
The interplay of the initial deformation of the PLF$^*$ caused by the reaction, 
Coulomb proximity, and the rotation of the 
PLF$^*$ results in the observed anisotropic angular distribution.
Changes in the shape of the
angular distribution with excitation energy are interpreted as 
being the result of forming more elongated initial geometries in the
more peripheral collisions.

\end{abstract}
\pacs{PACS number(s): 25.70.Mn} 

\maketitle

Emission of $\alpha$ particles and other light clusters from heavy nuclei is 
traditionally explained as an evaporative process governed by the available 
excitation energy. In contrast, understanding the fission of a heavy nucleus 
involves a description of the evolution of collective degrees-of-freedom (shape)
as the nuclear system re-organizes itself from a single nucleus into two heavy nuclei.
It has been proposed that thermal shape 
fluctuations, i.e. deformation, play a significant role in the statistical 
emission of large clusters \cite{Moretto75}. 
The importance of such thermal fluctuations has recently been realized 
within a multistep statistical model code \cite{Charity01}. 
One means of accessing large deformation  
is through the non-central collision of two heavy ions at 
intermediate energies (20$\leq$E/A$\leq$100 MeV). 
Following the exchange
of mass, charge, and energy between the projectile and target nuclei, an
excited projectile-like (PLF$^*$) and 
target-like fragment (TLF$^*$), which are deformed, are produced. 
As the two reaction partners separate they undergo decay
resulting in final residues referred to as the PLF and TLF. It is important to understand
how this decay is impacted by both the initial configuration of the nuclei and their
proximity to one another.

Dynamical fragment production, intermediate between the TLF and PLF, has been previously
 reported and has been associated with neck 
fragmentation of a transient nuclear system \cite{Casini93, Montoya94, Toke95}.
Recent analyses at intermediate energies \cite{Bocage00, Davin02, Piantelli02, Gingras02, Colin03}
have elucidated some of the essential characteristics of this decay mode.
These dynamical decays are clearly
distinguished from standard statistical decay of the 
undeformed PLF* and TLF*.
These processes \cite{Bocage00, Davin02, Colin03}  
can be viewed in a fluid dynamical perspective 
as the double neck rupture of a highly deformed dinuclear system 
\cite{Brosa90} and correspond to decay on a 
short timescale. 
It has been suggested that the initial configuration may not 
only be important for decay 
on a short timescale but may also impact fragment emission
on longer timescales \cite{Durand95, Gingras02}.
In this Rapid Communication, we present direct evidence on the importance of the 
persistence of the initial deformation and its interplay with Coulomb
proximity for longer-timescale fragment emission. 

In an experiment performed at Michigan State University, we bombarded a 
$^{92}$Mo target foil  5.45 mg/cm$^2$ thick with $^{114}$Cd ions accelerated to
E/A = 50 MeV by the K1200 cyclotron.  
A key aspect of this experiment was the detection of a PLF (15$\le$Z$\le$46) at
very forward angles (2.1$^{\circ}$$\le$$\theta_{lab}$$\le$4.2$^{\circ}$) with 
good angular ($\Delta$$\theta$=0.13$^\circ$,$\Delta$$\phi$=22.5$^\circ$) and energy 
resolution. Identification of the PLF by the 
$\Delta$E-E technique provided better than unit resolution, $\delta$Z/Z $\approx$0.25.  
Associated charged particles emitted at larger angles 
(7$^{\circ}$$\le$$\theta_{lab}$$\le$58$^{\circ}$) were measured in  
LASSA, a large area silicon strip array \cite{Davin01, Wagner01}.
The kinematic coverage in the experiment allowed us to reconstruct 
the decay of the PLF$^*$\cite{Yanez03}. 
The excitation of the PLF$^*$, which is  
correlated with its velocity damping, 
was deduced from
the measured multiplicities and 
kinetic energies of the detected particles in a calorimetric analysis \cite{Yanez03}.

\begin{figure}[t] 
\includegraphics [scale=0.40]{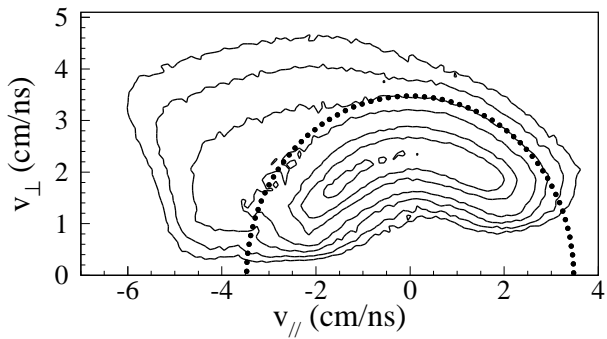}%
\caption{\label{fig:fig1}
Invariant cross-section map (linear scale) of $\alpha$ particles 
in the PLF$^*$ frame for the reaction $^{114}$Cd + $^{92}$Mo at E/A = 50 MeV.
The dotted line depicts the energy cut described in the text.
} 
\end{figure}

The binary nature of the collisions studied is presented in 
Fig.~\ref{fig:fig1}. Alpha particles 
detected in LASSA, coincident with a PLF, are shown in the 
invariant cross-section map presented in Fig.~\ref{fig:fig1}. The 
dominant feature evident in this figure is
an essentially circular ridge of yield centered on the reconstructed PLF$^*$ velocity 
with a radius that corresponds to the Coulomb repulsion between 
the emitted $\alpha$ and the PLF$^*$ residue, for a touching spheres configuration. 
In addition to standard 
evaporation, dynamical processes may also contribute to the yield along the 
Coulomb ridge. Also evident in Fig.~\ref{fig:fig1} is 
the contribution attributable to emission from a mid-velocity 
'source' \cite{Plagnol00,Gingras02}.
For the remainder of this work we focus on the yield along the Coulomb ridge,
which corresponds to a well-defined configuration between the 
emitted $\alpha$ and PLF$^*$ residue.

\begin{figure}[t!] 
\includegraphics [scale=0.35]{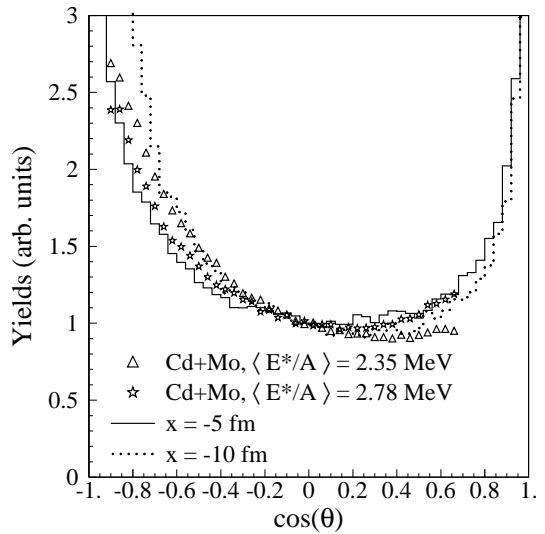}%
\caption{\label{fig:fig2}
Comparison of the angular distribution of $\alpha$ particles from the 
reaction $^{114}$Cd + $^{92}$Mo at E/A = 50 MeV with 
E$_{\alpha}$$\le$25 MeV in the PLF$^*$ frame with results of the
Langevin model.
} 
\end{figure}

In Fig.~\ref{fig:fig2} the yield of 
detected $\alpha$ particles with E$_{\alpha}$$\le$25 MeV (as indicated by the dotted 
line in Fig.~\ref{fig:fig1}), 
i.e. dominated by the ridge in Fig.~\ref{fig:fig1}, 
is displayed in the PLF$^*$ frame. 
To further define the events under investigation, we have selected on excitation energy 
($\langle$E$^*$/A$\rangle$$\approx$2.35 MeV and $\approx$2.78 MeV) 
by selecting on the velocity of the 
PLF$^*$\cite{Yanez03}. For these selected events, the average Z$_{PLF}$ 
detected is $\approx$34 ($\approx$32) at
$\langle$E$^*$/A$\rangle$=2.35 (2.78) MeV, while the reconstructed atomic number of the 
PLF$^*$ is on average $\approx$37 ($\approx$38).  The most notable 
feature in Fig.~\ref{fig:fig2} is the large forward-backward asymmetry 
present for the experimental data (open symbols). 
The data are backward peaked, towards the TLF, with an enhancement factor 
of $\approx$2.3-2.6 relative to cos($\theta$)=0. 
The finite acceptance of the experimental setup, evident in Fig.~\ref{fig:fig1},
limits the measured angular distribution to the range
-0.85$\leq$ cos($\theta$)$\leq$0.65.
The
lower excitation data manifests a broader angular distribution as compared to 
the higher excitation data. 
The lines displayed in this figure will be discussed later.

Large backward-forward asymmetries for intermediate mass fragments have
been previously observed and have been related to a dynamical fission-like process
\cite{Montoya94, Bocage00,Davin02,Colin03}. This process which is characterized
by highly aligned decay and relative velocities well above the Viola
fission systematics \cite{Viola85}, has been related to the neck rupture
of a deformed nuclear system. The large relative velocities observed for these aligned 
decays are believed to be intrinsically related to the collision dynamics. In
this work we focus on the asymmetry associated with a simpler configuration, 
namely one that results in fragments with Coulomb dominated kinetic energies. 
To understand the observed asymmetry associated with these simpler configurations
we compare the 
experimental data with the predictions of a one-dimensional Langevin model.

\begin{figure}[t] 
\includegraphics [scale=0.30]{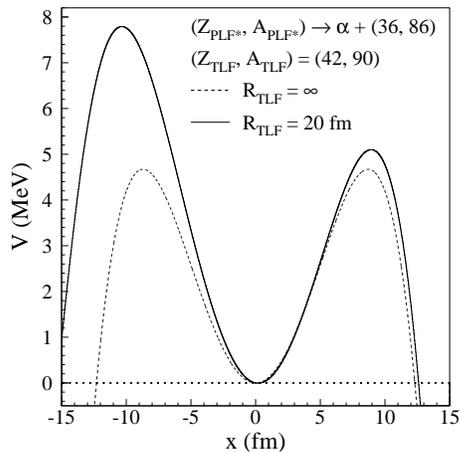}%
\caption{\label{fig:fig3}
One dimensional potential energy diagram of $\alpha$ particle emission from the
PLF* indicating the influence of the TLF$^*$ proximity. 
} 
\end{figure}

In this schematic model, we consider the emission of an $\alpha$ particle 
from a PLF$^*$ (Z=38,A=90) as it moves away from the TLF$^*$. The emission 
process of the $\alpha$ particle
from the PLF$^*$ is governed by a potential which simulates both the 
nuclear and Coulomb interaction between the $\alpha$ particle and the PLF$^*$ residue, as well
as their Coulomb interaction with the TLF$^*$.
This potential energy is parameterized by:
\begin{equation}
V(x) = -(x-c)(x+c)(\frac{x}{d})^2 + \sum_{i=1}^2{e^2\frac{(Z_{TLF*})(Z_i)}{R_i}} 
\end{equation}
where x is the separation distance between the $\alpha$ particle and the 
center-of-mass of the
$\alpha$-PLF$^*$ residue system. Thus, x is related to the 
configuration of the PLF$^*$ system and can be associated with its deformation. 
The constants c and d, which define the potential, 
are taken to have values of 12.3 and 35, respectively in order to determine the location
of the barriers. The detailed shape of the potential, other than the location of 
the barriers has been arbitarily chosen.
The second term in the potential
describes the interaction of the TLF* with the $\alpha$ particle and PLF$^*$ residue with 
R$_1$ and R$_2$ designating the relevant distances.  
For simplicity, we describe the influence of the TLF$^*$ at a distance
R$_{TLF}$ from the center-of-mass of the PLF$^*$ system i.e. R$_{TLF}$=R$_1$-x. 
Shown in Fig.~\ref{fig:fig3} is the relative potential normalized so that
V=0 at x=0. When the $\alpha$ particle-residue system decays in isolation 
(R$_{TLF}$= $\infty$), the 
potential is symmetric with a local minimum at x=0. Equal height 
barriers govern backward and forward emission as shown by the dashed line
in Fig.~\ref{fig:fig3}. 
The presence of the TLF$^*$ at a finite initial distance 
(e.g. 20 fm) from the $\alpha$-residue center-of-mass modifies the potential
through its Coulomb influence, 
increasing the 
backward barrier more than the forward barrier. A slight outward shift in the 
barrier positions is also evident in Fig.~\ref{fig:fig3} as indicated by the solid 
line. For a system 
with an initially spherical configuration (x=0) this asymmetry in the barriers favors
forward emission corresponding to a forward peaked angular distribution. 
The experimentally observed yield enhancement in the backward direction therefore 
suggests that the system is initially deformed towards the TLF$^*$--as one might expect 
from the nuclear collision process. As the TLF$^*$ and PLF$^*$ separate, the 
Coulomb proximity effect is diminished leading to equal height barriers 
for forward and backward emission. At a TLF$^*$-PLF$^*$ separation distance of 50 fm, the
two barriers are essentially equal corresponding to a sensitivity
for time up to $\approx$ 5 zs (1 zs = 1 x 10$^{-21}$s). 

To describe the evolution of an intial configuration as the PLF$^*$ moves away from the
TLF$^*$, the schematic model utilizes
a Langevin approach. 
The observed angular asymmetry suggests the persistence of the
initial configuration. 
This indicates that the motion is overdamped rather than underdamped, the latter 
would give rise to forward-backward oscillations of the configuration resulting in
both forward and backward peaking of the angular distribution.
Consequently, we work within the high friction limit which allows
us, in this schematic model, to eliminate inertial terms.
From an initial position on the potential, the change 
in position of the $\alpha$ particle is given by:

\begin{equation}
\Delta x=\frac{F\Delta t}{\beta} + k \cdot  \sqrt[]{\frac{2T\Delta t}{\beta}}
\end{equation}

\noindent where $\Delta$t is a time step of 0.1 zs. 
The first term describes the influence of the potential
on the particle's motion with the force due to the potential represented by F, while
$\beta$ corresponds to the friction. 
The impact of thermal motion on the particle's trajectory is included in the second term.
The fluctuating term, k, is taken to be a gaussian of unit width 
centered on zero with the 
magnitude of the thermal motion scaled for each time step by the 
temperature, T and the friction, $\beta$. The center-of-mass of the PLF$^*$ 
and its decay products is conserved at all times.

\begin{figure}[t] 
\includegraphics [scale=0.35]{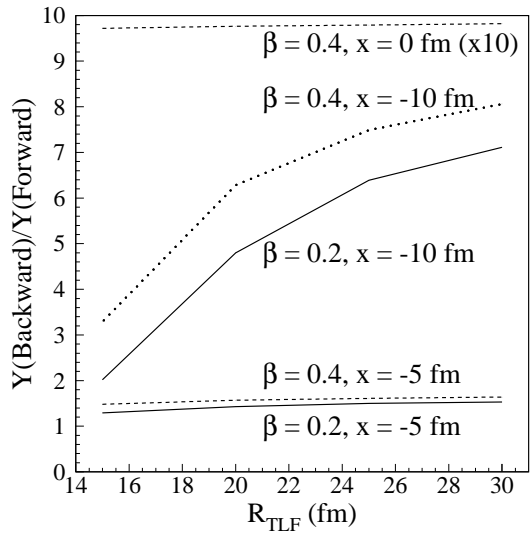}%
\caption{\label{fig:fig4}
Yield ratio for backward to forward emission as a function of the initial 
TLF$^*$-PLF$^*$ distance. Calculations for different initial configurations 
and friction are shown.
} 
\end{figure}

From its original position on the potential the position of 
the $\alpha$ is allowed to evolve in time. If the $\alpha$ particle surpasses either 
barrier the time is recorded. By calculating the fate of the
$\alpha$ particle for several initial cases, the relative yield of 
backward to forward emission, Y(backward)/Y(forward), is determined. 
Displayed in Fig.~\ref{fig:fig4}, is the dependence of the 
ratio Y(backward)/Y(forward) on the initial TLF$^*$-PLF$^*$ distance, 
R$_{TLF}$, for different conditions of initial configurations and friction. 
The temperature was assumed to be 4 MeV in all cases consistent with the re-constructed
excitation energy for the experimental data shown in Fig.~\ref{fig:fig2}.
For the case of an initially spherical configuration (x=0), the 
ratio Y(backward)/Y(forward), is slightly
less than unity and increases slightly with increasing R$_{TLF}$.  
This result is a direct manifestation of the preference for
forward emission due to the Coulomb proximity effect.
For a slightly elongated  initial configuration,
x = -5 fm, Y(backward)/Y(forward) 
exhibits a near constant value of 1.3-1.4. 
A larger value of the 
friction, namely $\beta$=0.4 (zs.MeV/fm$^2$), results in a marginally 
larger value for Y(backward)/Y(forward). For such initial configurations, the 
$\alpha$ particle is well inside both emission barriers. 
The independence of the yield ratio on R$_{TLF}$ can be understood  
if by the time the $\alpha$ particle reaches the top of the barrier, the 
TLF$^*$-PLF$^*$ distance is large and the proximity effect is small. 
Comparison of the x=0 and x=-5 fm cases clearly indicates that the magnitude of 
the ratio being larger than unity is related to the persistence of the 
initial configuration.
If the elongation associated with the initial configuration is large, namely 
comparable to the barrier position
of $\approx$ -10 fm, then Y(backward)/Y(forward) depends strongly on R$_{TLF}$. The
increase of this ratio with R$_{TLF}$ indicates that the time to 
overcome the backward barrier is sufficiently short for the Coulomb 
proximity of the TLF$^*$ to be important. 
For these elongated configurations, an enhanced 
sensitivity to the friction is also manifested. This enhanced sensitivity to the friction
can be understood by considering the motion of the $\alpha$ particle and the change of the
potential due to the proximity effect. For elongated configurations, 
namely a particle near the top of the left barrier, the gradient of the potential is small
resulting in the particle's motion being driven by the thermal term. In the case of larger
friction, $\beta$=0.4, the particle motion is reduced, hence the particle remains 
near the top of the barrier. As the TLF$^*$-PLF$^*$ separate, the importance of the 
Coulomb proximity is decreased leading to the decrease in magnitude of the barrier as 
well as a shift inwards leading to emission of the $\alpha$ particle. In the case of 
smaller 
friction, $\beta$=0.2, the particle initially at the top of the barrier has a 50$\%$ 
probability of moving inward sufficiently so as to be subsequently insensitive to the 
barrier changes induced by the Coulomb proximity effect.

The rotation of the decaying PLF$^*$ impacts the observed angular asymmetry. 
If the emission time is longer than a rotational period, a symmetric 
angular distribution would be observed independent of the initial configuration. 
The observed angular asymmetry 
signals that the emission time is short relative to a rotational period. 
Thus, the measured angular asymmetry presents a 'clock' 
which allows us to measure the time period between the initial separation of
the TLF$^*$-PLF$^*$ and the emission of the $\alpha$ particle. 
This scission-to-scission time has previously been 
investigated and found to depend systematically on the mass asymmetry of the PLF$^*$ 
decay \cite{Casini93}. For the largest mass asymmetries studied, 
times as short as 0.6 zs
were reported. The present case of $\alpha$ particle emission corresponds to a larger
asymmetry than previously investigated. 
If the $\alpha$ particle-PLF$^*$ residue constitutes  
a rotating di-nuclear system, its rotation resembles that of a classical macroscopic
object \cite{Kun91}. 
At lower bombarding energies this relationship between the
rotational frequency and the emission lifetime has been used to deduce the
fission timescale from the observed  
angular anisotropy for asymmetric fission \cite{Casini93}. 

For an assumed spin of the PLF$^*$, the emission time calculated in the model can be related
to an emission angle through the rotational frequency. 
We consider the rotation to be that of two spheres separated by the distance between the
$\alpha$ particle and the PLF$^*$ residue.
One means of ascertaining the spin of the PLF$^*$ is through the out-of-plane 
angular distribution. Unfortunately, limited out-of-plane kinematical coverage 
in this experiment prohibits 
us from using this approach. We have therefore assumed, consistent with 
the out-of-plane widths for similar experiments \cite{Casini93}, a spin of 40$\hbar$
which corresponds to a rotational period of 3.7 zs.
The angular distributions predicted for initial configurations of x=-5 fm and x=-10 fm are
displayed in Fig.~\ref{fig:fig2} as solid and dotted histograms respectively
for R$_{TLF}$=15 fm and $\beta$=0.4. To facilitate comparison 
of the shape of the predicted angular distribution with the experimentally 
measured angular
distribution, the distributions have been normalized to cos($\theta$)=0. 
The case of slightly elongated configurations results in a narrower 
angular distribution at backward angles. 
Under the assumption of small spin,
10 $\hbar$ (not shown), the angular distribution is more narrowly peaked at both
backward and forward angles, as expected. Moreover, it manifests a lesser 
dependence on the 
initial configuration. In this latter case it should be noted that the asymmetry in the 
total yield still depends strongly on the initial configuration as shown in 
Fig.~\ref{fig:fig4}.

Comparison of the model calculations with the experimental data reveals that 
under the assumption of a spin of 40$\hbar$ the lower excitation energy is better
described by x=-10 fm while the higher excitation is well described by x=-5 fm. 
In keeping with the schematic nature of the model calculation, no attempt was made
to reproduce the detailed shape of the measured angular distributions. 
As the higher excitation energy corresponds to 
larger velocity damping of the PLF$^*$, the observed trend suggests that more peripheral
collisons are associated with 
more elongated dinuclear
configurations i.e. more deformed geometries.

While this simple model can provide insight into the interplay
of deformation and Coulomb proximity effects in governing fragment emission, we emphasize 
that a more complete understanding will require a more realistic multi-dimensional 
model which includes a spin dependent potential, 
inertial terms \cite{Carjan86} and 
accounts for the presence of an 
initial kinetic energy in the emission direction \cite{Bredeweg02} due to  
incomplete velocity damping in the collision. In addition, the 
full de-excitation cascade should be accounted for. To allow investigation of 
small initial separations between 
the TLF$^*$ and PLF$^*$, nuclear proximity effects also need to be 
incorporated into the model.

Alpha decay of an excited PLF$^*$ following a peripheral
heavy ion collision at intermediate energy is examined. 
Emission in which the fragments manifest predominantly Coulomb kinetic energies exhibits 
a strongly anisotropic emission pattern, 
favoring emission toward the TLF$^*$. 
This enhancement of backward emission over forward emission can be 
related to the initial deformation of the PLF$^*$, induced by the 
nuclear interaction with the target, and can be affected by the Coulomb interaction 
between the separating TLF$^*$ and the PLF$^*$ system. 
For large initial deformation, the ratio of backward to forward emission manifests a 
large sensitivity to
the Coulomb proximity effect and to the magnitude of the nuclear friction. The
latter must be large in order for the motion to be overdamped.
Comparison of the experimental angular distribution with the model calculations indicates
that more peripheral collisions are associated with more elongated geometries. 
Although the nuclear interaction undoubtedly results in a distribution of 
initial configurations, the
simple model presented here captures the essence of the association between this initial 
configurational bias created by the reaction and the observed angular asymmetry.

\begin{acknowledgments}
	We would like to acknowledge the staff at MSU-NSCL for
providing the high quality beams which made this experiment possible. We gratefully 
acknowledge several stimulating conversations with A.S. Botvina on the topic of the
Coulomb proximity effect.  This work was supported by the
U.S. Department of Energy under DE-FG02-92ER40714 (IU), 
DE-FG02-87ER-40316 (WU) and the
National Science Foundation under Grant No. PHY-95-28844 (MSU).\par
\end{acknowledgments}

\bibliography{langevin.bib} 

\end{document}